\documentclass[11pt]{article}
\usepackage{amsmath,amssymb,color,graphics,epsfig,cite,subfloat,amsthm,graphicx,caption,
	epstopdf,float,subfig,subfloat,geometry,blindtext,indentfirst,cite,hyperref,cleveref,xcolor}

\usepackage{amsfonts}

\newcommand{\be}{\begin{equation}}
	\newcommand{\ee}{\end{equation}}
\newcommand{\bea}{\setlength\arraycolsep{2pt} \begin{eqnarray}}
	\newcommand{\eea}{\end{eqnarray}}

\def\fft#1#2{{\frac{#1}{#2}}}

\def\0{{\sst{(0)}}}
\def\1{{\sst{(1)}}}
\def\2{{\sst{(2)}}}
\def\3{{\sst{(3)}}}
\def\4{{\sst{(4)}}}
\def\5{{\sst{(5)}}}
\def\6{{\sst{(6)}}}
\def\7{{\sst{(7)}}}
\def\8{{\sst{(8)}}}
\def\sst#1{{\scriptscriptstyle #1}}

\def\vr{{\varPsi_{r}}}
\def\nr{{n^2+r^2}}

\begin{document}
	
	\begin{center}
		{\Large {\bf Scalarization of charged Taub-NUT black hole and the entropy bound }}
		
		\vspace{20pt}
		
		{\large Lei Zhang and Hai-Shan Liu}
		
		\vspace{10pt}

		\vspace{10pt}
		
		{\it Center for Joint Quantum Studies and Department of Physics,\\
			School of Science, Tianjin University, Tianjin 300350, China }
		
		\vspace{40pt}
		
		\underline{ABSTRACT}
		
	\end{center}

	We investigate the spontaneous scalarization of charged Taub-NUT black holes within the framework of Einstein-Maxwell-scalar-Gauss-Bonnet gravity. By selecting a suitable coupling function, the theory admits the analytic charged Taub-NUT geometry as a solution. We demonstrate that this scalar-free background becomes unstable within specific parameter regimes, leading to the bifurcation of a new branch of hairy charged Taub-NUT black holes. These solutions are characterized by a two-dimensional parameter space spanned by the electric charge and the NUT parameter. We conduct a systematic study of their properties, specifically the scalar charge, temperature, and entropy. Our analysis reveals that the entropy of the scalarized solutions exhibits particularly compelling features. Two universal characteristics emerge: first, the entropy of the hairy black hole is strictly greater than that of its scalar-free counterpart; second, the entropy reaches a local maximum precisely at the bifurcation point. Notably, when the electric charge is fixed, this maximum entropy value remains universal across a specific range of the mass parameter.

	\vfill{\footnotesize  leiz@tju.edu.cn ~~~~ hsliu.zju@gmail.com (Corresponding author)  }
	
	
	\thispagestyle{empty}
	\pagebreak
	
	

\newpage

\section{Introduction}
The black hole no-hair theorem is a foundational postulate in general relativity and black hole physics, emerging from a series of uniqueness theorems in the 1960s. It states that a stationary, asymptotically flat black hole is uniquely determined by only three macroscopic, observable parameters: mass, electric charge, and angular momentum \cite{Bekenstein:1972ny,Bekenstein:1995un}. Considerable effort has been dedicated to evaluating the validity of the black hole no-hair theorem. Subsequent research has demonstrated that the no-hair theorem may be circumvented by considering black holes interacting with various matter fields\cite{Volkov:1989fi,Bizon:1990sr,Greene:1992fw,Maeda:1993ap,Luckock:1986tr,Droz:1991cx}. In these scenarios, the black hole can support a non-trivial scalar field profile outside the event horizon. Hairy black hole solutions have also been successfully constructed within the framework of Horndeski theories, where the scalar field is kinetically coupled to the curvature\cite{Anabalon:2013oea,Sotiriou:2013qea, Sotiriou:2014pfa, Bergliaffa:2021diw}. Consequently, the thermodynamics of these configurations and their associated holographic applications have been intensively investigated \cite{Feng:2015oea,Feng:2015wvb,Jiang:2017imk,Liu:2018hzo,Hu:2018qsy}.

An alternative approach to generating black hole solutions with scalar hair involves the direct coupling of a scalar field to curvature invariants\cite{Pani:2011xm,Yunes:2011we,Pani:2011gy}. Within this context, extended scalar-tensor-Gauss-Bonnet (estGB) gravity has been investigated extensively in the literature\cite{Doneva:2017bvd,Antoniou:2017hxj,Silva:2017uqg,Minamitsuji:2018xde,Cunha:2019dwb,Brihaye:2018bgc}. For specific classes of coupling functions, the classical Schwarzschild black hole remains a valid solution of the theory. However, below a critical mass threshold, the Schwarzschild solution becomes unstable, giving rise to a new branch of scalarized black holes. This phenomenon is widely referred to as spontaneous scalarization. The generalization of this framework to incorporate electric charge was subsequently developed, extending the study of scalarization to the Reissner-Nordström(RN) black hole\cite{Herdeiro:2018wub,Doneva:2018rou,Wang:2020ohb,Jiang:2023gas}.

Beside RN black hole, the charged Taub-NUT black hole represents another fundamental solution to Einstein-Maxwell theory \cite{Plebanski:1975xfb}. In addition to mass ($m$) and electric charge ($q$), this solution is characterized by the NUT parameter ($n$), which imparts a gravomagnetic structure to the spacetime; notably, the RN solution is recovered in the limit where $n \to 0$. Recently, a wealth of  investigations have focused on interpreting the NUT parameter as an independent thermodynamic charge or 'hair' \cite{Hennigar:2019ive,Wu:2019pzr,Liu:2022wku,Yang:2023hll,Liu:2023uqf,Chen:2024knw}, analogous to the role of electric charge. Motivated by the status of the charged Taub-NUT geometry as a solution to Einstein-Maxwell theory, we investigate its spontaneous scalarization triggered by a scalar field coupled to the Gauss-Bonnet invariant. We have successfully constructed hairy charged Taub-NUT black hole solutions. Our results show that below a specific parameter threshold, the scalar-free background becomes unstable, leading to the bifurcation of a new branch of hairy solutions. These scalarized configurations are thermodynamically preferred, as they possess higher entropy than their bald counterparts. Furthermore, a novel phenomenon emerges: these hairy solutions exhibit a universal maximum entropy bound across a specific range of parameters.

The structure of this paper is organized as follows. In Section 2, we introduce the Einstein-Maxwell framework supplemented by a free scalar field coupled to the Gauss-Bonnet invariant. We adopt a specific coupling function that allows the theory to admit the analytic charged Taub-NUT black hole as a solution. A probe scalar analysis is then performed to identify the parameter space where spontaneous scalarization originates. In Section 3, we numerically construct scalarized charged Taub-NUT black holes across a wide range of parameters and conduct an intensive investigation of their physical and thermodynamic properties. Finally, we present our concluding remarks in Section 4.

\section{Einstein-Maxwell theory with a scalar coupled to Gauss-Bonnet terms }
In this section, we consider Einstein-Maxwell theory with a scalar non-mimimally coupled to Gauss-Bonnet combination, which is given by

\begin{equation}\label{action}
	S=\frac{1}{16\pi}\int d^4x \sqrt{-g}( R-\frac{1}{4} F_{\mu\nu} F^{\mu\nu}-2\nabla_{\mu}\varphi\nabla^{\mu}\varphi+\lambda^2\xi(\varphi)R_{GB}^2)
\end{equation}

where $R$ is the Ricci scalar with respect to the metric $g_{\mu\nu}$, $\varphi$ is the scalar field, $\xi(\varphi)$ is the coupling function with coupling constant $\lambda$ and $R_{GB}^2=R^2+R_{\mu\nu\rho\lambda}R^{\mu\nu\rho\lambda}-4R_{\mu\nu}R^{\mu\nu}$ is the Gauss-Bonnet invariant, $F_{\mu\nu}=(dA)_{\mu\nu}$ is the Maxwell tensor.

The equations of motion of the theory can be obtained  from the variation of the action(Eq. \ref{action}) with respect to the metric $g_{\mu\nu}$, the scalar field $\varphi$ and Maxwell field $A_\mu$,
\begin{equation}
	R_{\mu\nu}-\frac{1}{2}Rg_{\mu\nu}+\Gamma_{\mu\nu}=\frac{1}{2}(F_{\mu\rho}F^{\rho}_{\nu}-\frac{1}{4}F_{\alpha\beta} F^{\alpha\beta}g_{\mu\nu})+2\nabla_{\mu}\varphi\nabla_{\nu}\varphi-g_{\mu
		\nu}\nabla_{\alpha}\varphi\nabla^{\alpha}\varphi
\end{equation}

\begin{equation}
	\nabla_{\alpha}\nabla^{\alpha}\varphi=-\frac{\lambda^2}{4}\frac{d\xi(\varphi)}{d\varphi}R_{GB}^2
\end{equation}

\begin{equation}
	\partial^{\mu}(\sqrt{-g}F_{\mu\nu})=0
\end{equation}

Here $\nabla_{\mu}$ is the covariant derivative with respect to the metric $g_{\mu\nu}$ and the $\varGamma_{\mu\nu}$ is defined by

\begin{equation}
	\begin{split}
		\varGamma_{\mu\nu}=&-2R\nabla_{\mu}\varPsi_{\nu}-4\nabla^{\alpha}\varPsi_{\alpha}(R_{\mu\nu}-\frac{1}{2}Rg_{\mu\nu})+8R_{\mu\alpha}\nabla^{\alpha}\varPsi_{\nu}\\
		&-4g_{\mu\nu}R^{\alpha\beta}\nabla_{\alpha}\varPsi_{\beta}+4R^{\beta}_{\mu\alpha\nu}\nabla^{\alpha}\varPsi_{\beta}
		\,.\end{split}
\end{equation}

with $\varPsi_{\mu}$ given by

\begin{equation}
	\varPsi_{\mu}=\lambda^2\frac{d\xi(\varphi)}{d\varphi}\nabla_{\mu}\varphi  \,.
\end{equation}

The theory could possess quite different properties for different forms of coupling function $\xi(\varphi)$. In order for the theory can go back to Einstein-Maxwell theory when the scalar field $\varphi$ is absence, the coupling function should satisfy $\fft{\delta \xi}{\delta \varphi}|_{\varphi = 0} = 0$. It is obvious that the  Reissner-Nordstr\"{o}m  as well as charged Taub-NUT black holes with trivial scalar $\varphi=0$ are solutions of the theory.  Furthermore, if additional conditions  $\fft{\delta^2 \xi}{\delta \varphi ^2}|_{\varphi = 0} > 0$ is imposed, scalarized black holes could emerge \cite{Doneva:2017bvd,Liu:2024bzh}.

Various coupling functions have been studied in the literature\cite{Doneva:2017bvd,Antoniou:2017hxj,Brihaye:2018bgc}, we choose the exponential coupling function

\begin{equation}
	\xi\left(\varphi\right)=\frac{1}{12}\left[1-\exp(-\alpha\varphi^2)\right] \,.
\end{equation}	

It satisfies $\fft{\delta \xi}{\delta \varphi}|_{\varphi = 0} = 0$ and $\fft{\delta^2 \xi}{\delta \varphi ^2}|_{\varphi = 0} \ne 0$, a tachyonic instability could arise which leads to spontaneous scalarization or normal scalarization.

The scalariztion of RN black hole has been intensely studied\cite{Herdeiro:2018wub,Doneva:2018rou}.  In this work, we are interested in charged Taub-NUT black holes, which is given by	

\begin{equation}
	ds^{2}=-f\left(dt+2n\cos\theta d\phi\right)^{2}+\frac{dr^2}{h}+\left(r^2+n^2\right)\left(d\theta^2+\sin^2\theta d\phi^2\right) \,
\end{equation}

with

\begin{equation}
	h=f=\frac{r^2-2mr-n^2+q^2}{r^2+n^2},
\end{equation}

\begin{equation}
	A=\frac{qr}{r^2+n^2}(dt+2ncos\theta d\phi).
\end{equation}

The solution contains three integration constants, $m$ and $q$ are the mass and electric charge parameters and $n$ is the NUT parameter. 	

Before constructing the scalarized black hole solution, we shall first consider a test scalar field $\delta \varphi$ under the charged Taub-NUT background, and the linear equation of motion of test scalar is given by

\begin{equation}
	\nabla_{\alpha}\nabla^{\alpha}\delta\varphi+\frac{\lambda^2}{4}\delta\varphi R_{GB}^2=0 \,,
\end{equation}

and the Gauss-Bonnet combination $R_{GB}^2$ on the charged Taub-NUT background is

\begin{align}
	R_{GB}^2=\frac{8}{\left(n^2+r^2\right)^6}&\left\{6n^8+5 q^4 r^4-6 n^6 \left(2 q^2+15 r^2\right)+5 n^4 \left(q^4+24 q^2 r^2+18 r^4\right)\notag\right.
	\\
	\phantom{=\;\;}
	&\left.-6 m^2 \left(n^6-15 n^4 r^2+15 n^2 r^4-r^6\right)-2 n^2 \left(19 q^4 r^2+30 q^2 r^4+3 r^6\right)\notag\right.
	\\
	\phantom{=\;\;}
	&\left.12mr\left[6n^6 - q^2 r^4 - 5 n^4 \left(q^2 + 4 r^2\right) +
	2 n^2 \left(5 q^2 r^2 + 3 r^4\right)\right]\right\}\,.
\end{align}

Performing a standard spherical harmonics decomposition of the scalar field\cite{Herdeiro:2018wub}
$\delta\varphi\left(r,\theta,\phi\right)=\sum_{lm}Y_{lm}\left(\theta,\phi\right)U_l
\left(r\right)$, the radial equation of the scalar turns to a Schrodinger-like equation
\begin{equation}\label{scalar equation}
	\frac{1}{\left(r^2+n^2\right)}\frac{d}{dr}\left[\left(r^2-2mr-n^2+q^2\right)
	\frac{dU_l}{dr}\right]-\left[\frac{l\left(l+1\right)}{n^2+r^2}-\frac{\lambda^2}{4}R_{GB}^2\right]U_l=0 \,.
\end{equation}	

It now becomes an eigenvalue problem under appropriate boundary conditions, e.g. fixing $(n\,, \lambda\,, q)$ the equation can admit solutions for a discrete mass parameter $m$. It means a perturbed mode could exist which may imply a black hole solution with scalar hair could bifurcate from the charged Taub-NUT black hole.

We only consider the zero mode $l=0$. Requiring the probe scalar field is regular on the horizon,  a Taylor expansion of scalar field near the black hole event horizon $r_h$ takes the form
\begin{equation}
	U_l\left(r\right)=u_0+ u_1\left(r-r_h\right)+{\cal O} (r-r_h)^2 + \dots \,.
\end{equation}

and the equation of motion constrains the coefficient  $u_1$ to be
\begin{equation}
	u_1=\frac{ u_0\left[3n^6+n^4\left(3r_h^2-6 q^2\right)+3n^2\left(q^4-r_h^4\right)
		-r_h^2 \left(q^4-6q^2r_h^2+3r_h^4\right)\right]\lambda^2}{r_h\left(n^2+r_h^2\right)^3\left(n^2-q^2+r_h^2\right)} \,.
\end{equation}
\begin{figure}[H]
	\centering
	\subfloat[Fixed $q/\lambda=1/4$]{
		\centering
		\includegraphics[width=0.7\linewidth]{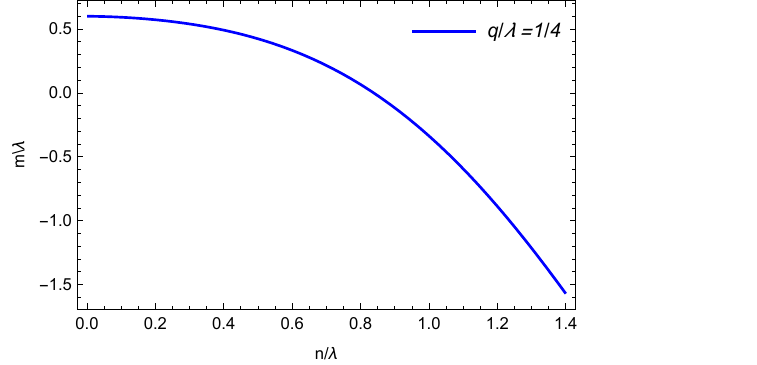}	
	}
	\\
	\subfloat[Fixed $n/\lambda$]{
		\centering
		\includegraphics[width=0.5\linewidth]{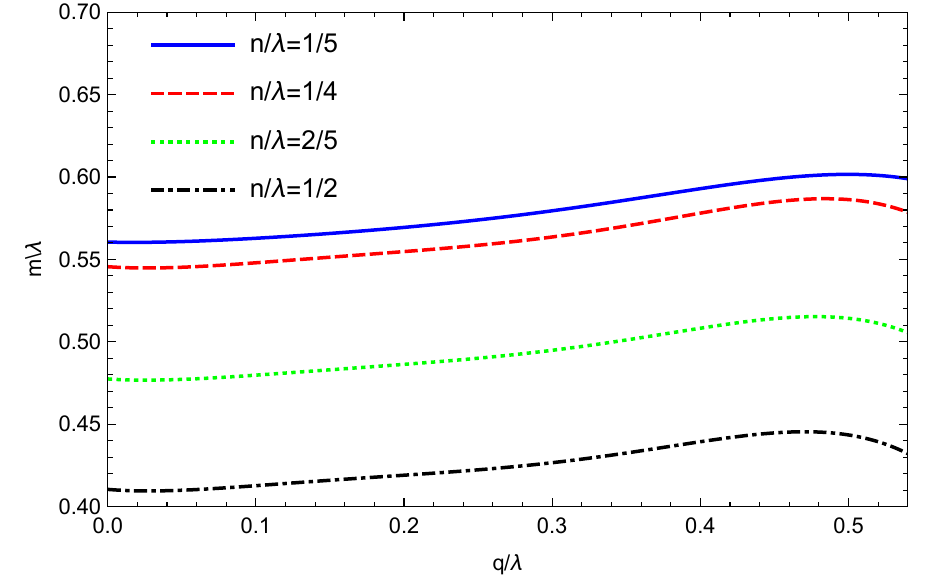}	
	}
	\subfloat[Fixed $n=q$]{
		\includegraphics[width=0.5\linewidth]{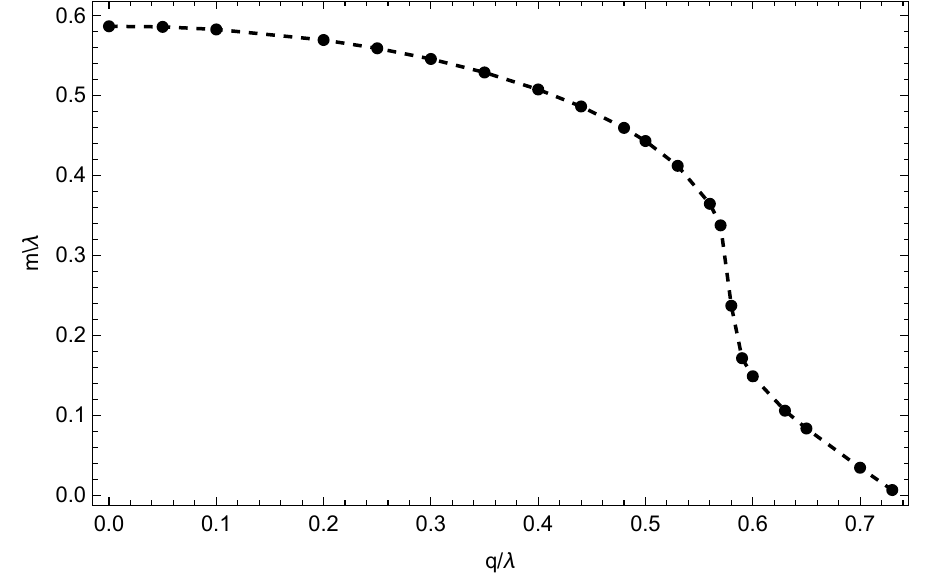}
	}
	\caption{The existence line for fixed $q/\lambda\,$, fixed $ n/\lambda \,$ and $ n = q$.}
\end{figure}

Though the  probe scalar equation is linear, it  can not be solved analytically, however it can be solved numerically.  Given the value of $(n\,,\lambda\,, q)$, one can numerically obtain the value of $m$. One can use the coupling constant $\lambda$ to make other parameters be dimensionless $(q/\lambda) \,, n/\lambda \,, m/\lambda$, and the perturbation equation gives a constraint on these parameters. Here, we display three cases fixed $q/\lambda$, fixed $n/\lambda$ and $n=q$. The existence curves for different values of $q$ are similar, thus we selected a single representative value of $q/\lambda =1/4$ as an example.	 For different values of $n/\lambda$, the  $m(q)$ curves are far from each other, but they have the similar pattern.  As showing in Fig.1,  with the increases of $q/\lambda$, the value of $m/\lambda$ will increase first to a maximum value and  then decrease at some ending point. And ending point means that we can't find a proper solution for the scalar equation(Eq. \eqref{scalar equation}).  As for the case of fixed $n=q$, the  $m(q)$ curve is a monotone decreasing function. The horizon radius $r_h$ of the analytical solution reduces to $r_h = 2m$ when $n = q$, and since a negative $r_h$ is unphysical, the curve must therefore end at $m=0$.

\section{Numerical results}
In this section, we shall construct charged Taub-NUT black holes with a non-trivial scalar hair. Since the equations of motion are difficult to solve, we shall turn to numerical methods. We consider the Taub-NUT-like metric ansatz 
\begin{equation}
	ds^2=-f(r)(dt+2n cos\theta d\phi)^2+\frac{dr^2}{h(r)}+(r^2+n^2)(d\theta+sin^2\theta d\phi^2)
\end{equation}
with $h(r)$ and $f(r)$ are undetermined functions of radial coordinate $r$.
And vector potential has the same form with that of analytical charged Taub-NUT solution
\begin{equation}
	A=V(r)(dt+2ncos\theta d\phi) \,,
\end{equation}
with $V(r)$ is a function of $r$.
The scalar field is static and spherical $\varphi$=$\varphi(r)$. Substituting these into the equations of motion, we can obtain five field equations.
	\begin{flalign}\label{eom1}
		&\
		f\left(n^2+r^2\right)\left\{4n^2V^2+\left(n^2+r^2\right)\left[h^{'}\left(r+2\lambda^2\varPsi_{r}\right)-1\right]\right\}-4\lambda^2h^2f\left[2n^2\varPsi_{r}+r\left(n^2+r^2\right)\varPsi_{r}^{'}\right]
		-&\nonumber
		\\
		&\
		4\lambda^2h^3f\left(n^2+r^2\right)\left[2n^2+r^2+\left(n^2+r^2\right)^2\varphi^2-6r^2\lambda^2h\varPsi_{r}+4n^2\lambda^2\varPsi_{r}^{'}+4r^2\lambda^2\varPsi_{r}^{'}\right]+&\nonumber
		\\
		&\
		\left(n^2+r^2\right)^3hV^{'2}+3n^2\left(n^2+r^2\right)f^2\left(2\lambda^2h^{'}\varPsi_{r}+4\lambda^2h\varPsi_{r}^{'}-1\right)=0,&
	\end{flalign}

	\begin{flalign}
		&\
		\left(n^2+r^2\right)h\left\{\left(n^2+r^2\right)V^{'2}+f^{'}\left[r\left(n^2+r^2\right)+2\lambda^2\left(n^2+r^2-3r^2h\right)\varPsi_{r}\right]\right\}-&\nonumber
		\\
		&\
		\left(n^2+r^2\right)f\left\{n^2+r^2-4n^2V^2+h\left[-r^2+\left(n^2+r^2\right)^2\varphi^{'2}-6n^2\lambda^2f^{'}\varPsi_{r}\right]\right\}-&\nonumber
		\\
		&\
		n^2f\left(n^2+r^2+8r\lambda^2h\varPsi_{r}\right)=0,&
	\end{flalign}

	\begin{flalign}
		&\
		f\left\{\left(\nr\right)^2f^{'}h^{'}-2\left(\nr\right)h\left[rf^{'}\left(6\lambda^2h^{'}\vr-1\right)+\left(\nr\right)\left(2V^{'2}-f^{''}\right)\right]\right\}-&\nonumber
		\\
		&\
		\left(\nr\right)hf^{'2}\left(n^2+r^2-4r\lambda^2h\vr\right)-4n^2f^3\left(2\lambda^2h^{'}\vr+4\lambda^2h\vr^{'}-1\right)+&\nonumber
		\\
		&\
		2f^2\left\{-8n^2V^2+r\left(n^2+r^2\right)h^{'}+2h\left[n^2+\left(n^2+r^2\right)^2\varphi^{'2}\right]\right\}-&\nonumber
		\\
		&\
		8\lambda^2h^2f\left\{r\left(\nr\right)\vr f^{''}+f^{'}\left[n^2\vr+r\left(\nr\right)\vr^{'}\right]\right\}=0,&
	\end{flalign}

	\begin{flalign}
		&\
		fh(n^2+r^2)^4\varphi^{'}f^{'}+f^2\left(n^2+r^2\right)^4\varphi^{'}h^{'}+2f^2h\left(n^2+r^2\right)^2\left[2r(n^2+r^2)\varphi^{'}+(n^2+r^2)^2\varphi^{''}\right]+&\nonumber
		\\
		&\
		\left\{fh(n^2+r^2)^2\left[3r^2\lambda^2h^{'}f^{'}-2\lambda^2(n^2+r^2)f^{''}\right]-3n^2\lambda^2ff^{'}\left(n^2+r^2\right)^2\left(f^{'}h+fh^{'}\right)+\right.&\nonumber
		\\
		&\
		\phantom{=\;\;}
		\left.4n^2\lambda^2f^3\left[2h\left(n^2-3r^2\right)+r\left(n^2+r^2\right)h^{'}\right]+\lambda^2h(n^2+r^2)(n^2+r^2-r^2h)f^{'}+\right.&\nonumber
		\\
		&\
		\phantom{=\;\;}
		\left.2\lambda^2fh^2r(n^2+r^2)\left[2n^2f^{'}+r(n^2+r^2)f^{''}\right]-\lambda^2f\left(n^2+r^2\right)^3f^{'}h^{'}+\right.&\nonumber
		\\
		&\
		\phantom{=\:\:}
		\left.24n^2rf^2h(n^2+r^2)^2f^{'}-6n^2\lambda^2f^2h\left(n^2+r^2\right)^2f^{''}\right\}\frac{d\xi(\varphi)}{d\varphi}=0,&
	\end{flalign}

	\begin{flalign}\label{eom5}
		&\
		8n^2f^2V+\left(\nr\right)^2hf^{'}V^{'}+&\nonumber
		\\
		&\
		\left(\nr\right)f\left\{\left(\nr\right)h^{'}V^{'}+2h\left[2rV^{'}+\left(\nr\right)V^{''}\right]\right\}=0,&
	\end{flalign}
where $"'"$ is derivative with respect to radial coordinate $r$ where $\varPsi_r$ is defined by
	\begin{equation}
		\varPsi_r=\lambda^2\frac{d\xi(\varphi)}{d\varphi}\frac{d\varphi}{dr}  \,.
	\end{equation}

Taking a close look at the horizon $r_h$, requiring the metric functions $h$ and $f$ vanish simultaneously and imposing regularity conditions for $\varphi$, one can have 
	\begin{equation}
		f\left(r\right)|_{r\rightarrow\\r_h}\rightarrow 0,\quad
		h\left(r\right)|_{r\rightarrow\\r_h}\rightarrow 0,\quad
		\varphi\left(r\right)|_{r\rightarrow\\rh}\rightarrow \varphi_h,  \quad
		V\left(r\right)|_{r\rightarrow\\r_h}\rightarrow \psi_{e}  \,.
	\end{equation}
		Taking Taylor expansion near the event horizon $r_h$, the equations of motion tell us that the fields have the form
	\begin{align}
		&\
		h\left(r\right)=\frac{1+\delta}{r_h} \left(r-r_h\right)+O(r-r_h)^2  \,,
		\\
		&f\left(r\right)=f_1\left(r-r_{h}\right)+O(r-r_h)^2   \,,
		\\
		&\varphi\left(r\right)=\varphi_{h}+\frac{e^{6\varphi_{h}^2}\varphi_{h}\left[3n^2 \left(1+\delta\right)f_1-\lambda^2\left(3+6\delta+\delta^2\right)\right]r_{h}}{\left(1+\delta\right)\left[e^{12\varphi_{h}^2}\left(n^2+r_{h}^2\right)^2-4\lambda^4\varphi_{h}^2\delta\right]}\left(r-r_h\right)+O(r-r_h)^2 \,,
		\\
		&V\left(r\right)=\psi_{e}+\frac{f1r_{h}\left[e^{12\varphi_{h}^2}r_{h}\left(n^2+r_{h}^2\right)^2A-2\varphi_{h}^2B\right]}{\left(n^2+r_{h}^2\right)\sqrt{\left\{\left(1+\delta\right)\left[4\lambda^4\varphi_{h}^2\delta-e^{12\varphi_{h}^2}\left(n^2+r_{h}^2\right)^2\right]r_{h}\right\}}}\left(r-r_h\right)+O(r-r_h)^2 \,.
		\end{align}
		where $A$ and $B$ are
		\begin{align}
			&\
			A=\left[r_{h}^2\delta+n^2\left(4\psi_{e}^2+\delta\right)\right],
			\\
			&B=\left[-3f_{1}n^2\left(n^2+r_{h}^2\right)\left(1+\delta\right)+3r_{h}^3\left(1+\delta\right)^2+n^2r_{h}\left(3+6\delta+8\psi_{e}^2\delta+3\delta^2\right)\right]\lambda^4.
		\end{align}
There are four parameters on the horizon, $r_h$, $\delta$, $\varphi_h$ and $\psi_{e}$. $\psi_{e}$ is a parameter related to electric charge, $f_1$, $\delta$ are parameters charaterize the gravity mode, and $\varphi_h$ is parameter associated with scalar field. It is worth pointing out that  the effect of the Gauss-Bonnet coupling term appears in the leading term of the fields.

In the infinity, we take a large $r$ expansion and let the fields be	
	\begin{align}
		&\
		h\left(r\right)=h_0+\frac{h_1}{r}+\frac{h_2}{r^2}+\frac{h_3}{r^3}+\frac{h_4}{r^4}+\cdots,
		\\
		&f\left(r\right)=f_0+\frac{f_1}{r}+\frac{f_2}{r^2}+\frac{f_3}{r^3}+\frac{f_4}{r^4}+\cdots,
		\\
		&\varphi\left(r\right)=\varphi_0+\frac{\varphi_1}{r}+\frac{\varphi_2}{r^2}+\frac{\varphi_3}{r^3}+\frac{\varphi_4}{r^4}+\cdots,
		\\
		&V\left(r\right)=v_0+\frac{v_1}{r}+\frac{v_2}{r^2}+\frac{v_3}{r^3}+\frac{v_4}{r^4}+\cdots.
	\end{align}
	
Substituting these into the field equations(Eq. \eqref{eom1}-Eq. \eqref{eom5}), and solving the equations of motion order by order in $1/r$, one can find that the coefficients $(h_i\,,f_i\,,\varphi_i\,,v_i)$ can be can be expressed in terms of $(f_1\,,\varphi_1\,,v_1)$ for $i\geq2$. The parameters $f_1$, $\varphi_1$, and $v_1$ correspond to the physical properties of the system: black hole mass, scalar charge, and electric charge, respectively. Consequently, we redefine them as $f_1 = -2m$, $\varphi_1 = D$, and $v_1 = q$. For the subsequent construction of numerical solutions, we impose the boundary conditions in the infinity $v_0 = 0$, $\varphi_0 = 0$, and $f_0 = 1$.
	\begin{align}
		&\
		h\left(r\right)=1-\frac{2m}{r}+\frac{D^2+2\left(q^2-2n^2\right)+2 n^2-q^2}{r^2}+\frac{-4mn^2+m\left(D^2+6n^2\right)}{r^3}
		\\
		&\quad\quad\quad+\frac{6n^2-3n^2q^2+D^2\left[(4m^2+2n^2-q^2\left(1+\lambda^2\right)\right]}{3r^4}+\cdots,\nonumber
		\\
		&f\left(r\right)=1-\frac{2m}{r}-\frac{2n^2-q^2}{r^2}+\frac{m\left(D^2+6n^2\right)}{3r^3}
		\\
		&\quad\quad\quad+\frac{24n^4-12n^2q^2+D^2\left[8m^2+8n^2-q^2\left(4+\lambda^2\right)\right]}{12r^4}+\cdots,\nonumber
		\\
		&\varphi\left(r\right)=\frac{D}{r}+\frac{Dm}{r^2}-\frac{D\left[2D^2-16m^2+8n^2+6\left(q^2-2n^2\right)+q^2\left(\lambda^2-2\right)\right]}{12r^3}
		\\
		&\quad\quad\quad+\frac{D\left\{24m^3-2m\left(D^2+6n^2\right)-2m\left[3D^2-12n^2+q^2\left(\lambda^2+6\right)\right]\right\}}{12r^4}+\cdots,\nonumber
		\\
		&V\left(r\right)=\frac{q}{r}-\frac{q\left[6n^2+D^2\left(\lambda^2+1\right)\right]}{6r^3}-\frac{D^2mq\left(3\lambda^2+4\right)}{12r^4}+\cdots.\nonumber
		\end{align}

We are now in the stage of performing numerical calculations, and we have successfully obtained hairy charged Taub-NUT black hole solutions. We choose  coupling potential $\xi\left(\varphi\right)=\frac{1}{12}\left[1-\exp(-\alpha\varphi^2)\right]$, with $\alpha =6$. The resulting solutions are characterized by two primary parameters: the electric charge $q$ and the NUT parameter $n$. Together, these span a two-dimensional parameter space. We categorize our investigation into three trajectories, fixed $q$, fixed $n$ and the diagonal $q = n$, to provide a comprehensive map of the black hole's configurations. 
 
\subsection{Fixed q}
The numerically derived charged Taub-NUT solution admits a non-trivial scalar field. Here, we plot the scalar charge parameter $D$ as a function of the mass parameter $m$ for a range of NUT parameters $n$, with fixed $q/\lambda=1/4$ in Fig.2. The $Z_2$ symmetry inherent to the scalar field is clearly exhibited in these plots. Notably, the $D$-$m$ curves cross the horizontal axis at $D = 0$, indicating that the solution at these crossing points corresponds to a trivial scalar field ($\varphi = 0$). This, in turn, corresponds to the bifurcation point mentioned in the preceding section, where the hairy charged black hole branches off from the analytic scalar-free charged Taub-NUT black hole. The scalar charge parameter $D$ starts at zero at the bifurcation point and terminates at a finite value for a specific mass parameter $m$. As illustrated, an increase in the NUT parameter $n$ results in a decrease in the mass parameter $m$ at the bifurcation points, a trend consistent with the results presented in Fig. 1. Simultaneously, it leads to an increase in the absolute value of the scalar charge $D$ at the endpoint.Besides, a distinct feature emerges for the black dashed curve ($n/\lambda = 3/5$): as the mass parameter $m$ approaches zero, the deviation from the scalar-free solution grows sharply.
	 
\begin{figure}[H]
	\centering
	\subfloat[]{
		\centering
		\includegraphics[width=0.5\linewidth]{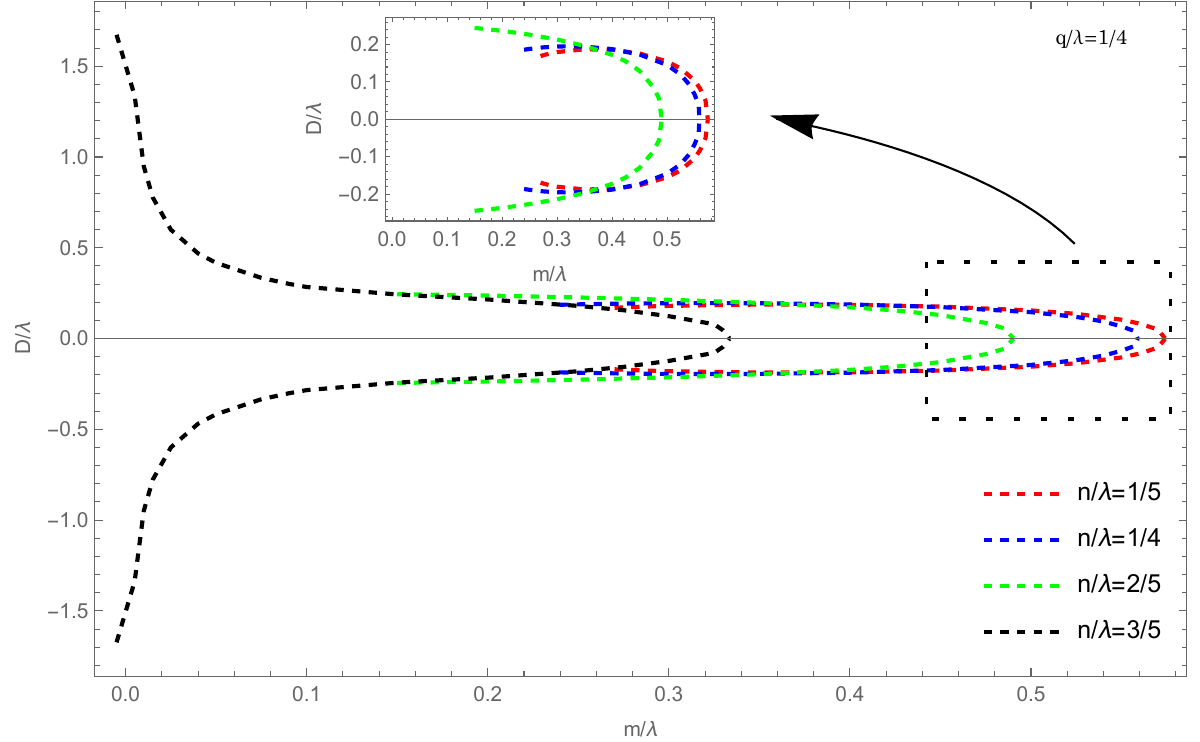}	
	}
	\subfloat[]{
		\centering
		\includegraphics[width=0.5\linewidth]{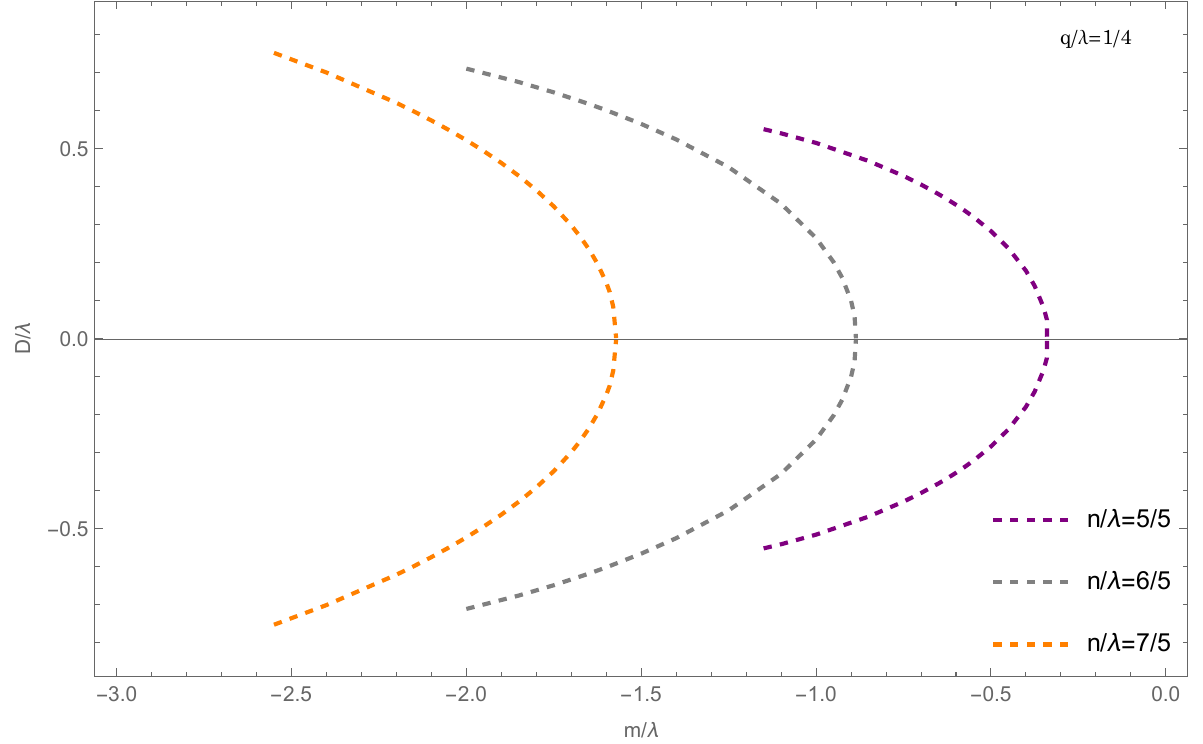}
	}
	\caption{The scalar charge parameter $D$ of the hairy charged Taub-NUT  black hole as a function of its mass parameter $m$, with the ratio $q/\lambda$ fixed at $1/4$ and various values of the NUT parameter $n$. The $D$-$m$ curves for which the value of $m$ at the bifurcation point is greater than zero are shown in the left panel (a), while those for which this value is less than zero are presented in the right panel (b).}
\end{figure}

Next, we plot the scalar charge $D$ of the black hole as a function of its event horizon radius $r_h$. This analysis reveals a novel characteristic: for an intermediate range of the NUT parameter $n$, two distinct solution branches can exist. The left panel of Fig. 3 displays the $D$-$r_h$ curves corresponding to small values of the NUT parameter, where only a single solution branch is observed for all cases. In contrast, the right panel presents the $D$-$r_h$ curves for large NUT parameter values $n$. As illustrated, for $n/\lambda = 1$ and $n/\lambda = 6/5$, two solution branches emerge. However, for slightly larger values of $n$, such as $n/\lambda = 7/5$, the solution reverts to a single branch once again.
\begin{figure}[H]
	\centering
	\subfloat[]{
		\centering
		\includegraphics[width=0.5\linewidth]{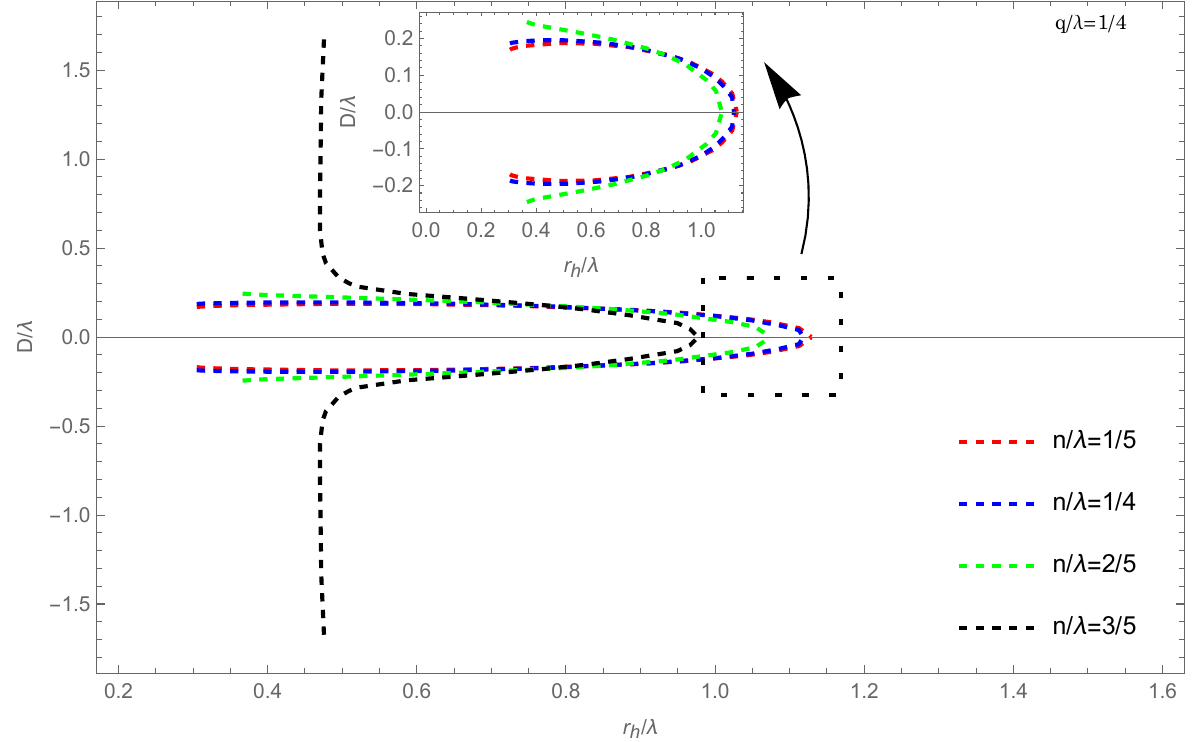}	
	}
	\subfloat[]{
		\centering
		\includegraphics[width=0.5\linewidth]{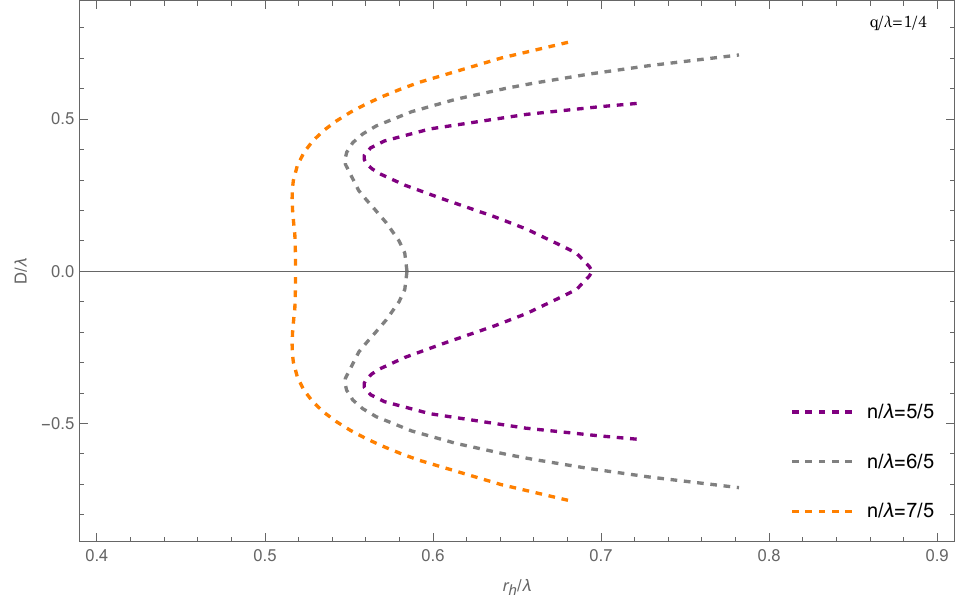}
	}
	\caption{The scalar charge parameter $D$ of the black hole is plotted as a function of its event horizon radius $r_h$. The $D$-$r_h$ curves corresponding to small values of the NUT parameter $n$ are shown in panel (a), where all curves exhibit only a single branch; by contrast, those for larger values of $n$ are presented in panel (b).}
\end{figure}

We now turn to the thermodynamic properties of the hairy black holes, focusing on the entropy and Hawking temperature. Owing to the presence of the higher-curvature Gauss–Bonnet term in the action, the entropy of these black holes no longer equals one-quarter of the event horizon area $A_h$. The correct entropy can be derived via the Wald entropy formula\cite{Wald:1993nt,Iyer:1994ys}

\begin{equation}
	S_H=-\frac{1}{8}\int_{r_h}\sqrt{h}dx^2 \epsilon_{ab}\epsilon_{cd} \frac{\partial L}{\partial R_{abcd}} \,,
\end{equation}
where $L$ denotes the Lagrangian and $\epsilon_{ab}$ is the volume form on the 2-dimensional cross-section of the horizon at $r_h$. For the extended scalar-tensor-Gauss–Bonnet theory considered herein, the explicit expression for the black hole entropy is given by

\begin{equation}
	S_H=\pi(r_h^2+n^2)+4\pi\lambda^2\xi\left(\varphi_h\right) \,,
\end{equation}
where $\varphi_h$ denotes the scalar field evaluated at the event horizon $r_h$, and $F(\varphi_h)$ is the coupling function of the theory.

\begin{figure}[H]
	\centering
	\subfloat[]{
		\centering
		\includegraphics[width=0.5\linewidth]{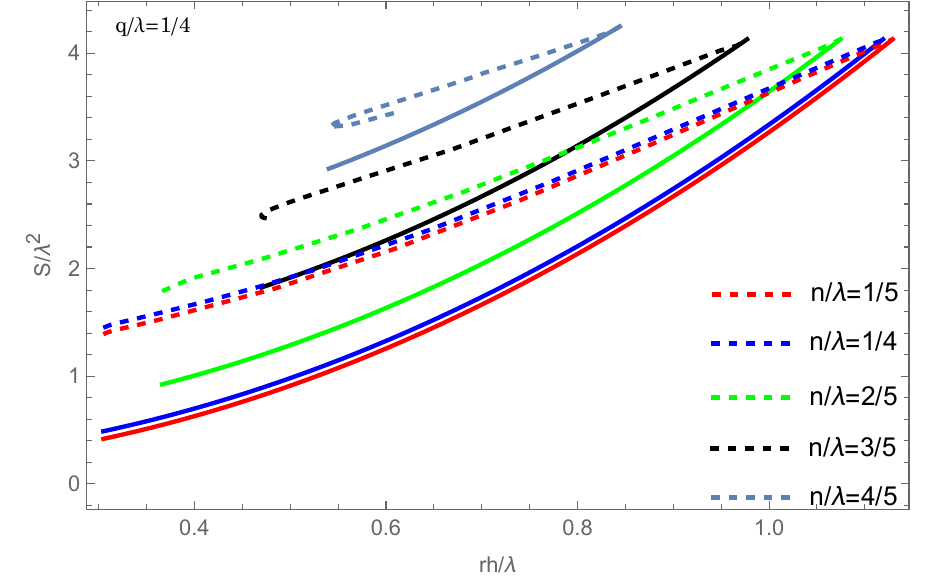}	
	}
	\subfloat[]{
		\includegraphics[width=0.5\linewidth]{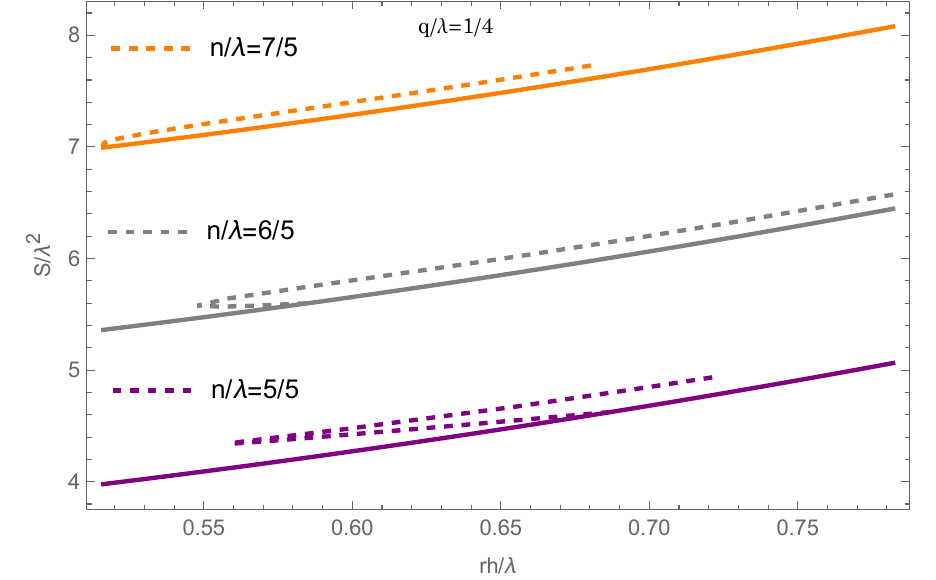}
	}
	
	\caption{$S/\lambda^2$ as a function of $r_h/\lambda$ for fixed $q/\lambda = 1/4$. Dashed lines denote charged scalarized black holes, while solid lines denote scalar-free Taub–NUT black holes for the same NUT parameter $n$.}
\end{figure}

Evidently, the entropy generally does not equal one-quarter of the event horizon area, but reduces to this standard value when the Gauss–Bonnet term is turned off by setting $\lambda = 0$. We plot the entropy $S$ as a function of the horizon radius $r_h/\lambda$ for fixed $q/\lambda = 1/4$ in panels (a) and (b) of Fig. 4. The dashed curves correspond to scalarized black holes, while the solid curves represent scalar-free charged Taub–NUT black holes with the same NUT parameter $n$. The entropy of scalarized black holes is consistently larger than that of their scalar-free counterparts, implying that the former are thermodynamically more stable and thus more favorable. The phenomenon of two solution branches existing for large values of the NUT parameter $n$ is clearly exhibited in panel (b).

As seen in panel (a) of Fig. 4, for a fixed NUT parameter $n$, the entropy of the hairy black hole exhibits a maximum at the bifurcation point, with these maximum values being nearly identical across different values of $n$. This phenomenon was also observed in the uncharged case ($q=0$) in \cite{Liu:2024bzh}.  To further clarify this phenomenon, we consider two additional values of the ratio $q/\lambda$, namely $2/5$ and $1/2$. Together with $q/\lambda = 1/4$, we plot the entropy as a function of the mass parameter for these three cases in Fig. 5. The entropy values at the bifurcation point remain constant over a certain range of the mass parameter, which is consistent with observations in the uncharged system. The influence of parameter $q$ manifests in two key aspects: first, it determines the specific constant value of the entropy, which decreases monotonically with $q/\lambda$; second, it modifies the valid range of the mass parameter within which the entropy retains this constant value. For instance, in the uncharged system, the maximum value of $m$ is nearly zero.

\begin{figure}[H]
	\centering
	\subfloat[$q/\lambda=1/4$]{
		\centering
		\includegraphics[width=0.65\linewidth]{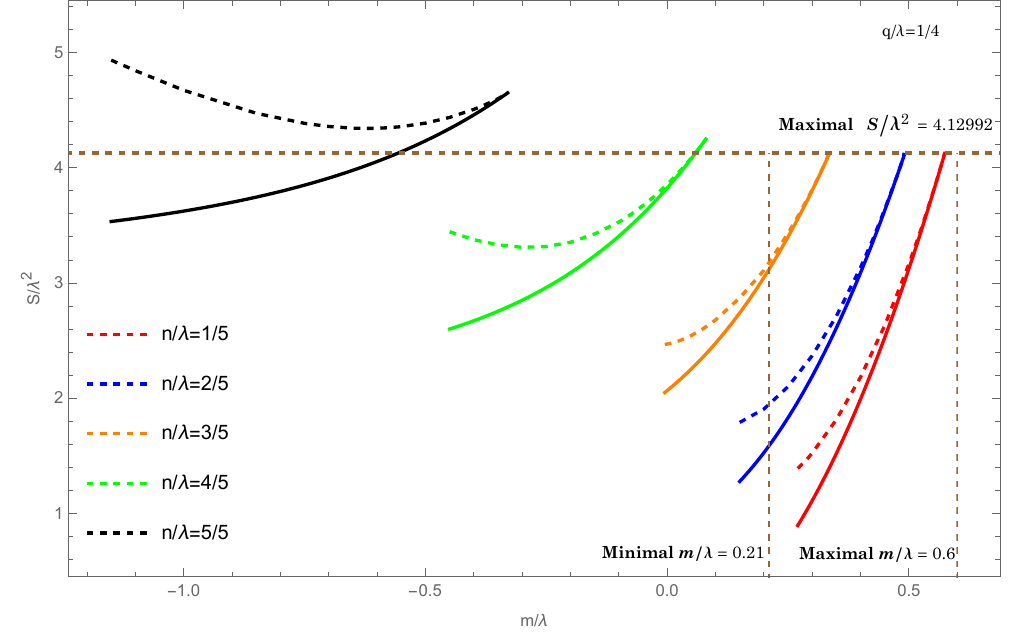}	
	}\\
	\subfloat[$q/\lambda=2/5$]{
		\includegraphics[width=0.5\linewidth]{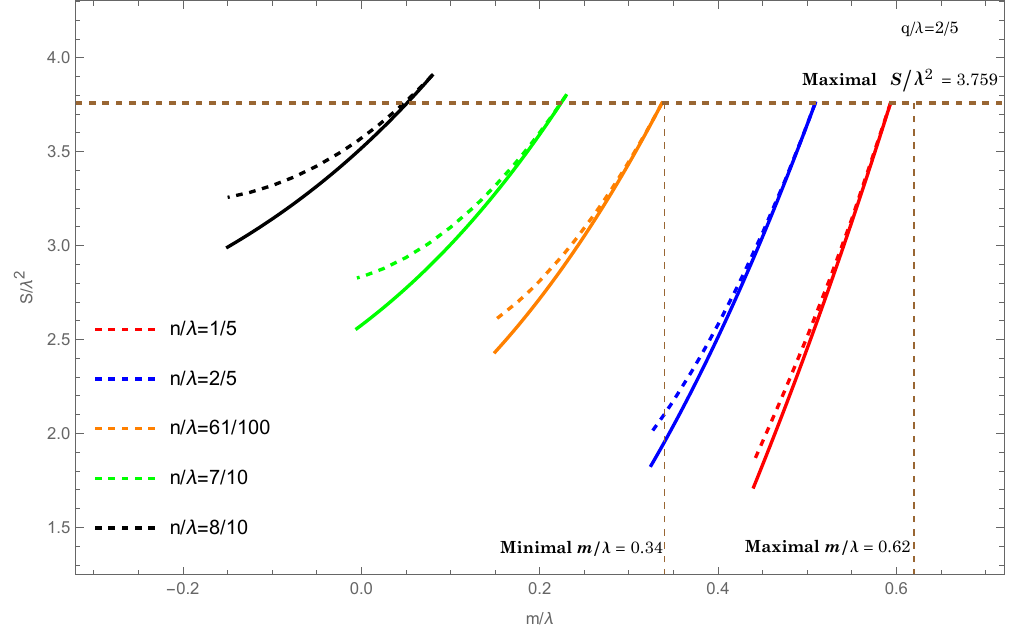}
	}
	\subfloat[$q/\lambda=1/2$]{
		\centering
		\includegraphics[width=0.5\linewidth]{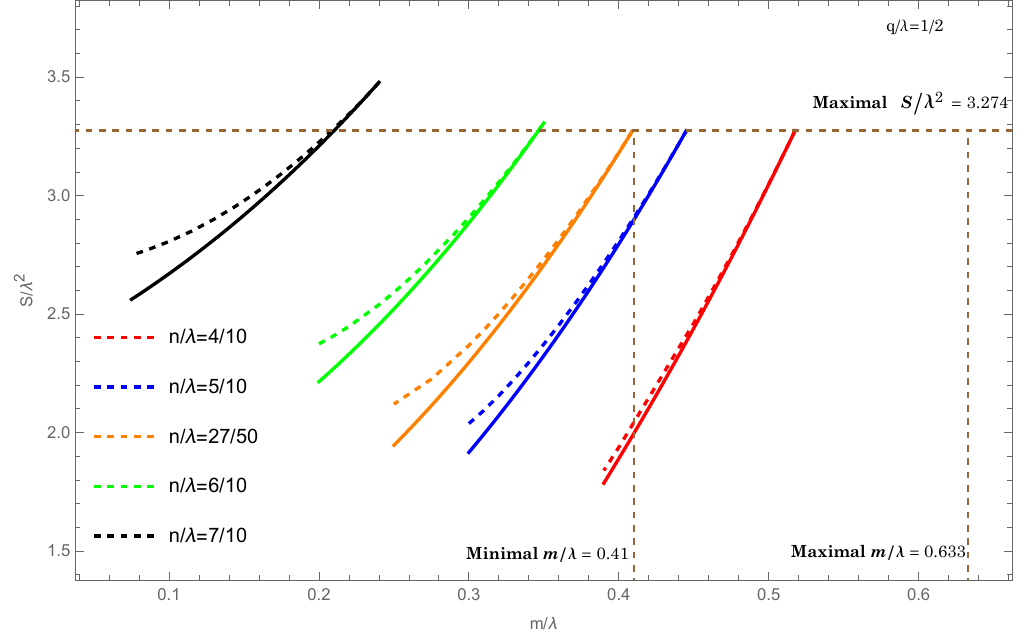}	
	}
\caption{$S/\lambda^2$ as a function of $m/\lambda$ for different values of the charge parameter. Dashed lines and solid lines correspond to charged scalarized black holes and scalar-free Taub–NUT black holes with the same NUT parameter $n$, respectively.}
\end{figure}

The entropy bound exhibits two primary features: a uniform maximum value across configurations, and the location of this maximum at the bifurcation points. Building on these observations, we plot two curves in Fig. 6(a): the constant entropy line and the existence line (defined by the collection of bifurcation points). It is evident that these two lines overlap across a specific range of the mass parameter. Mathematically, these lines can be represented as functions $n(m)$. To further analyze this, we plot the difference between these two functions, $\Delta n$, as a function of $m$ in Fig. 6(b), including a horizontal zero line for comparison. The results confirm that $\Delta n$ remains zero over the same mass parameter range. Finally, in Fig. 6(c–f), we present cases for $q/\lambda = 2/5$ and $q/\lambda = 1/2$, both of which demonstrate that the universal entropy bound exists within a certain range of the mass parameter.

\begin{figure}[H]
	\centering
	\subfloat[$q/\lambda=1/4$]{
		\centering
		\includegraphics[width=0.5\linewidth]{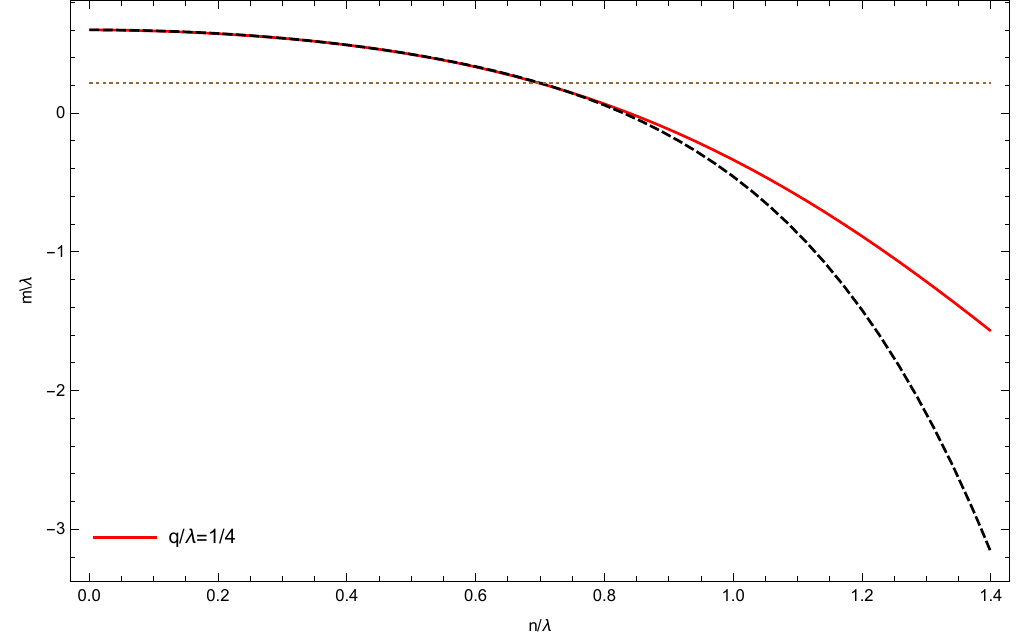}	
	}
	\subfloat[$q/\lambda=1/4$]{
		\includegraphics[width=0.5\linewidth]{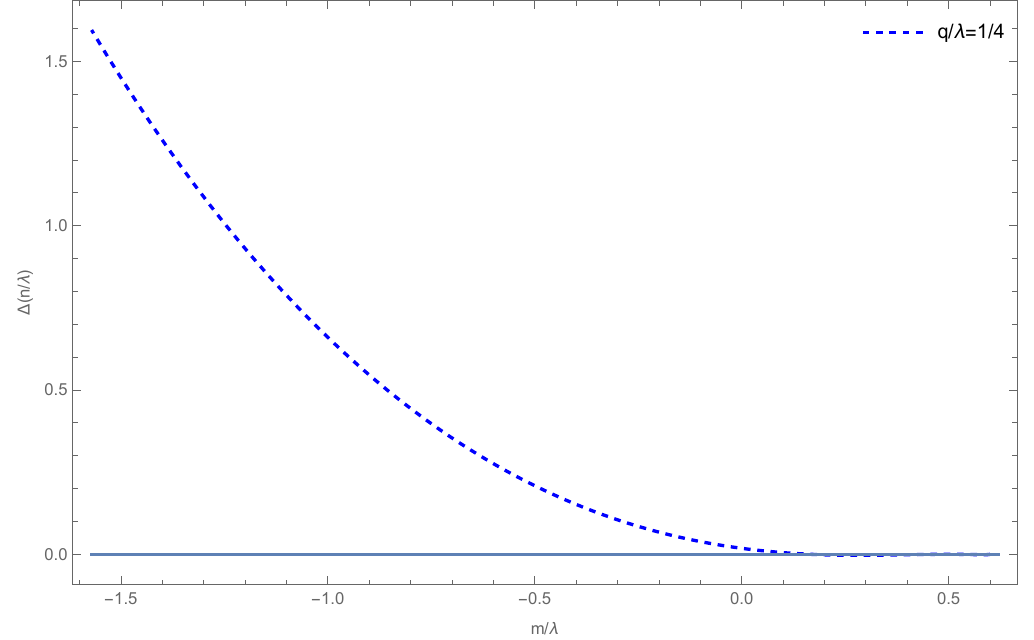}
	}
	\\
	\subfloat[$q/\lambda=2/5$]{
		\centering
		\includegraphics[width=0.5\linewidth]{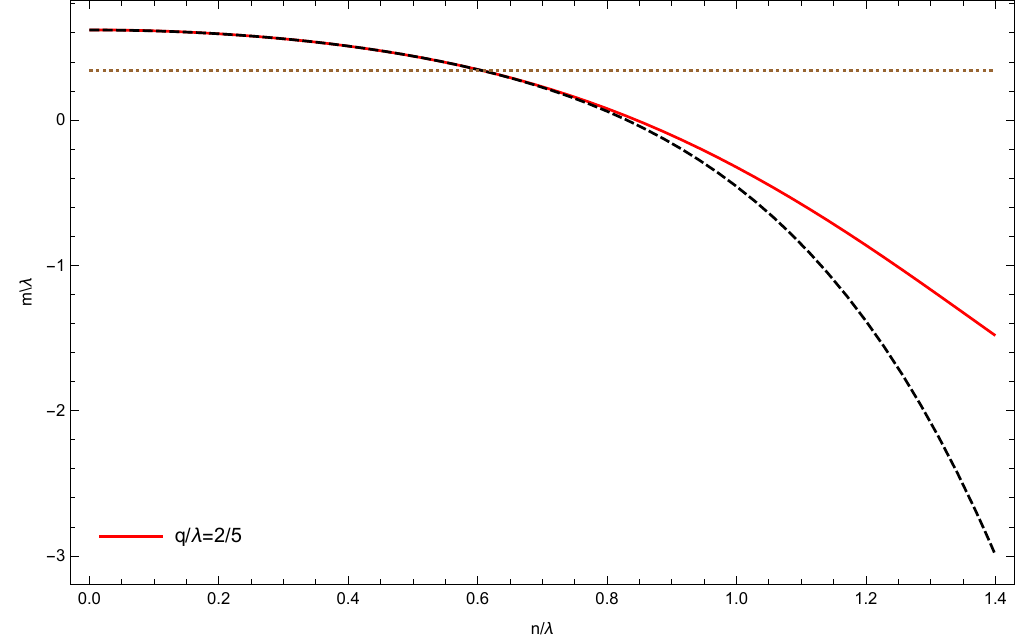}	
	}
	\subfloat[$q/\lambda=2/5$]{
		\includegraphics[width=0.5\linewidth]{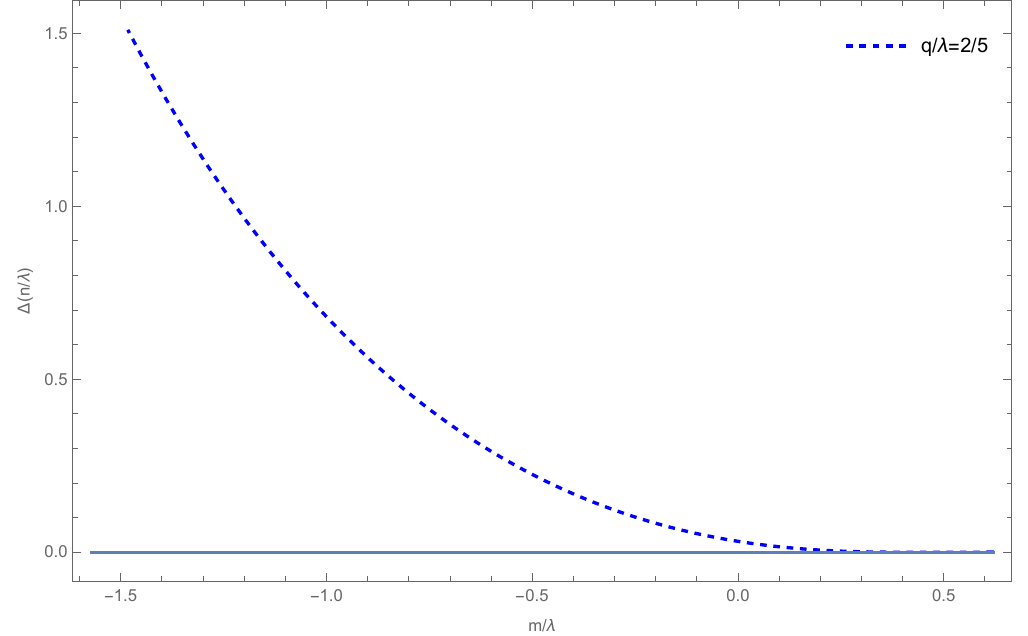}
	}
	\\
		\subfloat[$q/\lambda=1/2$]{
		\centering
		\includegraphics[width=0.5\linewidth]{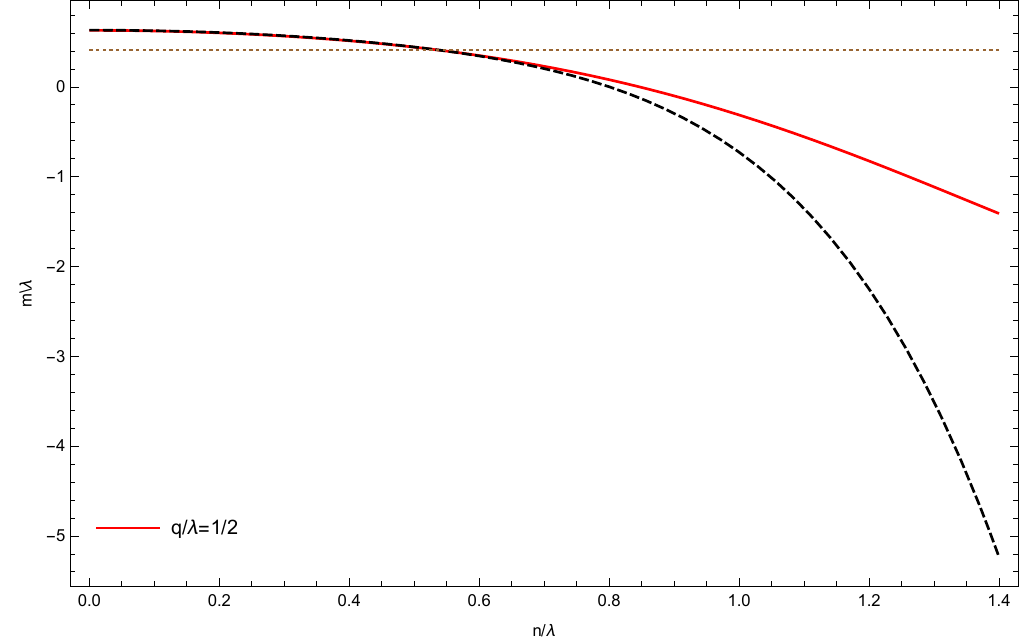}	
	}
	\subfloat[$q/\lambda=1/2$]{
		\includegraphics[width=0.5\linewidth]{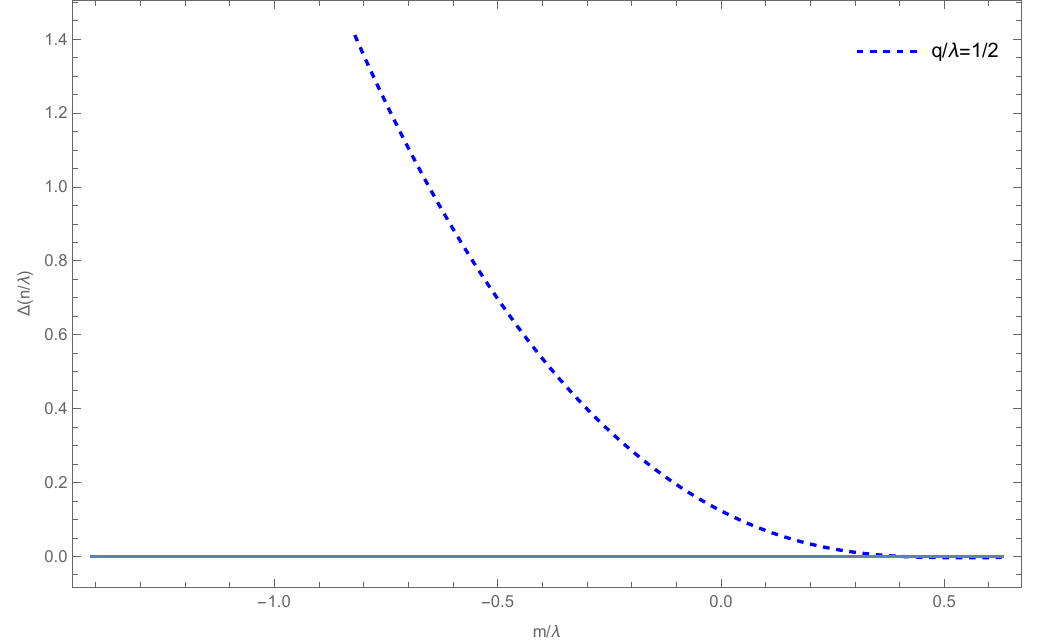}
	}
	\caption{The top-left panel (a) illustrates the existence line and the constant entropy line, while the top-right panel (b) depicts the difference between these two lines for $q/\lambda = 1/4$. Both panels demonstrate that a universal entropy bound exists across a specific range of the mass parameter. Similarly, panels (c, d) and (e, f) present analogous results for $q/\lambda = 2/5$ and $q/\lambda = 1/2$, respectively. }
	\end{figure}

As seen in Fig. 6, the range of the mass parameter over which the universal entropy bound emerges can be readily identified. For instance, the mass range is $(0.21, 0.6)$ for $q/\lambda = 1/4$ and $(0.41, 0.63)$ for $q/\lambda = 1/2$. This shift implies that increasing $q/\lambda$ constricts the mass range of the universal entropy bound. Finally, including the neutral case, the mass ranges for various values of $q/\lambda$ are plotted in Fig. 7.

\begin{figure}[H]
	\centering
		\includegraphics[width=0.8\linewidth]{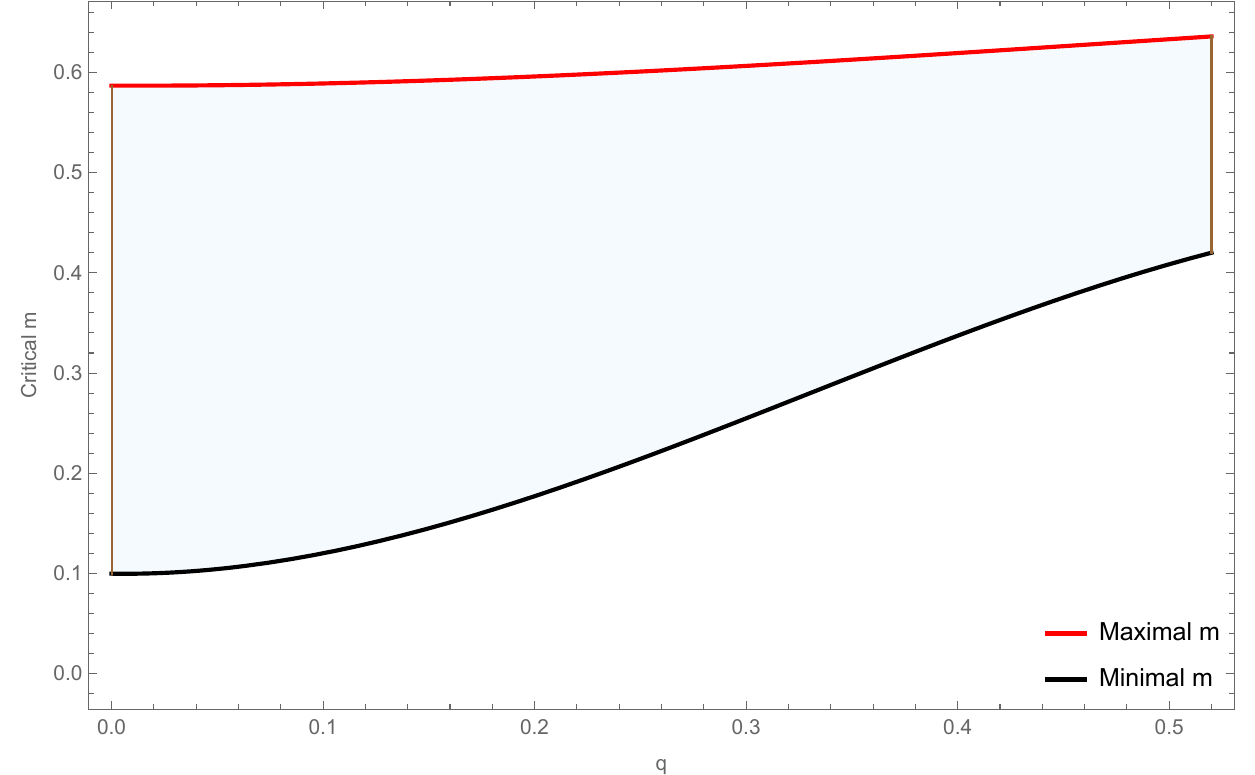}	
	\caption{The mass parameter ranges of the universal entropy bound for various values of $q/\lambda$. }
\end{figure}

The temperature of the charged hairy black holes can be calculated via the standard method. For our Taub–NUT-like metric, it is given by
\begin{eqnarray}
	T&=&\frac{\sqrt{h'(r_h) f'(r_h)}}{4\pi} \,.
\end{eqnarray}	
The temperature is plotted as a function of the mass parameter and the event horizon radius for various values of the NUT parameter $n$ and a fixed ratio $q/\lambda =1/4$ in Fig. 6. As can be seen, the temperature of the hairy black holes is consistently higher than that of their scalar-free counterparts.
\begin{figure}[H]
	\centering
		\includegraphics[width=0.8\linewidth]{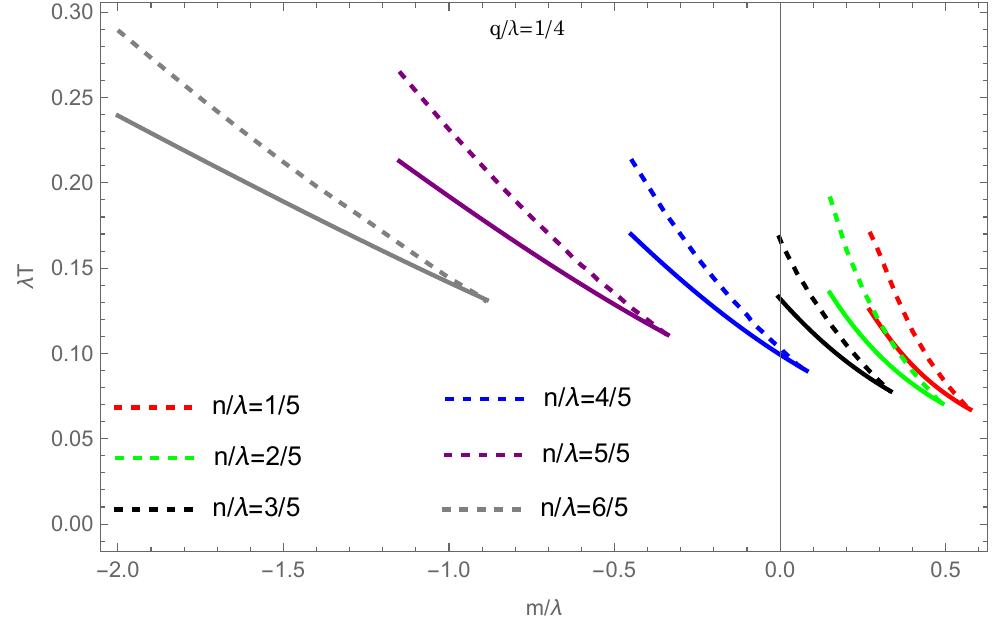}	
	\caption{$\lambda T$ as a function of $m/\lambda$. The dashed line and solid line correspond to the charged scalarized black hole and the scalar-free Taub-NUT black hole of the same $n$ respectively.}
\end{figure}

\subsection{Fixed $n$ and $n=q$}

\begin{figure}[H]
	\centering
	\subfloat[Fixed $n/\lambda=1/4$]{
		\centering
		\includegraphics[width=0.5\linewidth]{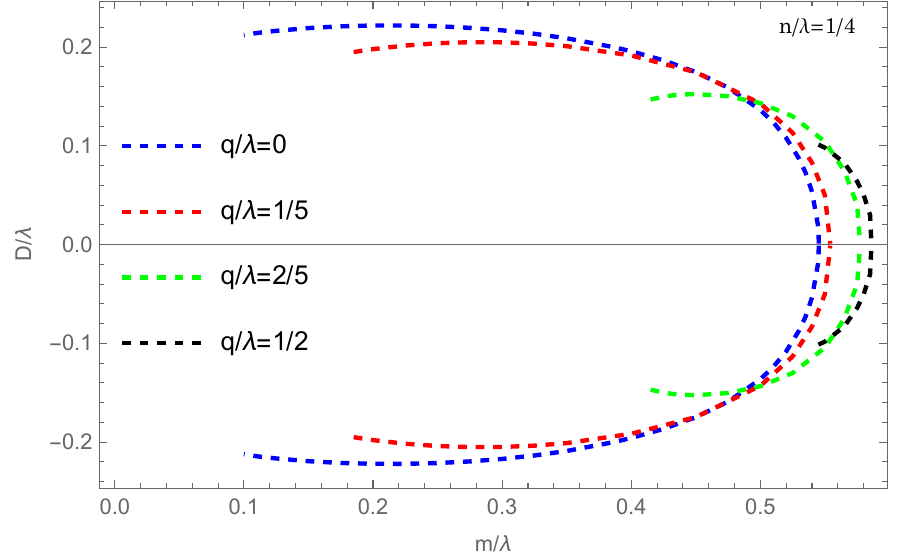}	
	}
	\subfloat[Fixed $n=q$]{
		\includegraphics[width=0.5\linewidth]{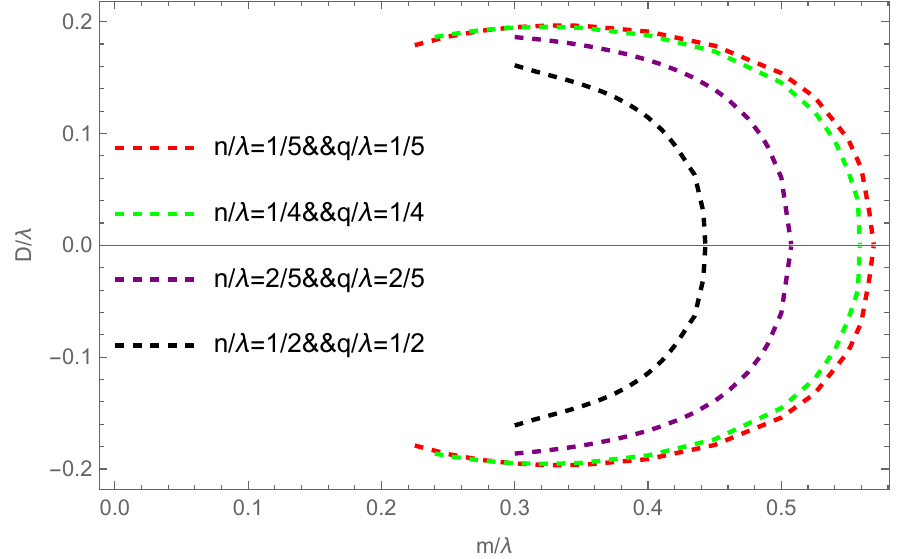}
	}
	\caption {Scalar charge parameter D as a function of the mass parameter $m$ for hairy black holes. Panel (a) displays the scalarized solutions with fixed ratio $n/\lambda=1/4$, while panel (b) corresponds to the case of $n=q$.}
\end{figure}

In the previous subsection, we intensively studied hairy black holes for fixed $q/\lambda$; we shall consider fixed NUT parameter $n$ and the case of $n=q$. We have already shown the bifurcation points of fixed $n$ and $n=q$ in Fig.1. The hairy charged black holes for these two cases can be obtained through the similar routine. In Fig.9, we plot the scalar charge parameter $D$ as a function of mass parameter $m$, with panel (a) showing the scalarized solutions for a fixed ratio $n/\lambda = 1/4$  and panel (b) for case of  $n = q$. 

The results for the uncharged case $(q=0)$ with $n/\lambda=1/4$ are also included in panel (a) for comparison. The influence of the electric charge on the scalarized solutions is evident. In contrast to the fixed-$q/\lambda$ case, for the fixed-$n/\lambda$ case, increasing $q/\lambda$ leads to both a decrease in the absolute values of $D/\lambda$ at the endpoint and a reduction in the curve lengths. For fixed $n=q$, a larger value of $n/\lambda$ (or $q/\lambda$) also yields a shorter curve.

\begin{figure}[H]
	\centering
	\subfloat[$S-m$ curve for fixed $n/\lambda=1/4$]{
		\centering
		\includegraphics[width=0.5\linewidth]{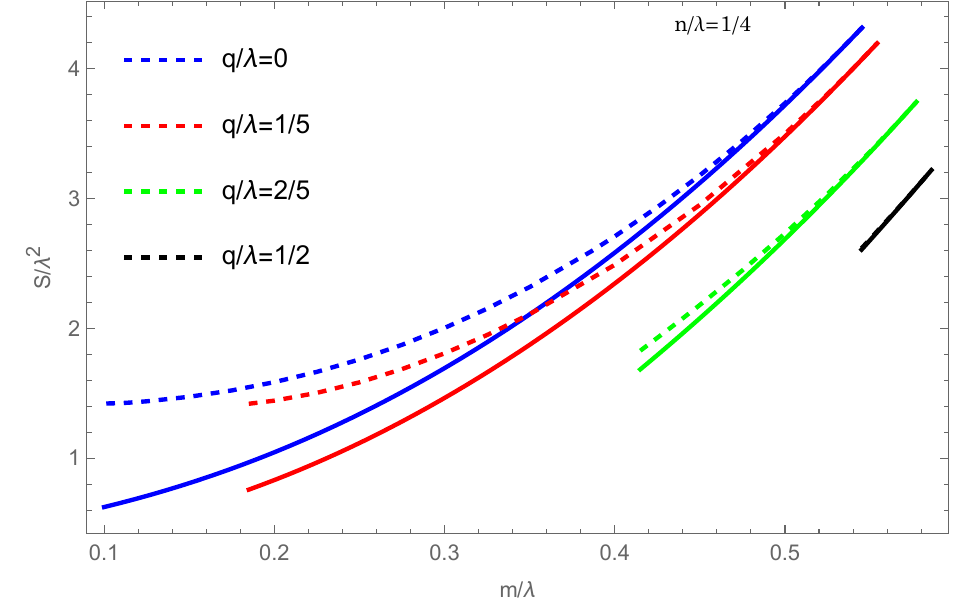}	
	}
	\subfloat[$S-m$ curve for for $n=q$]{
		\includegraphics[width=0.5\linewidth]{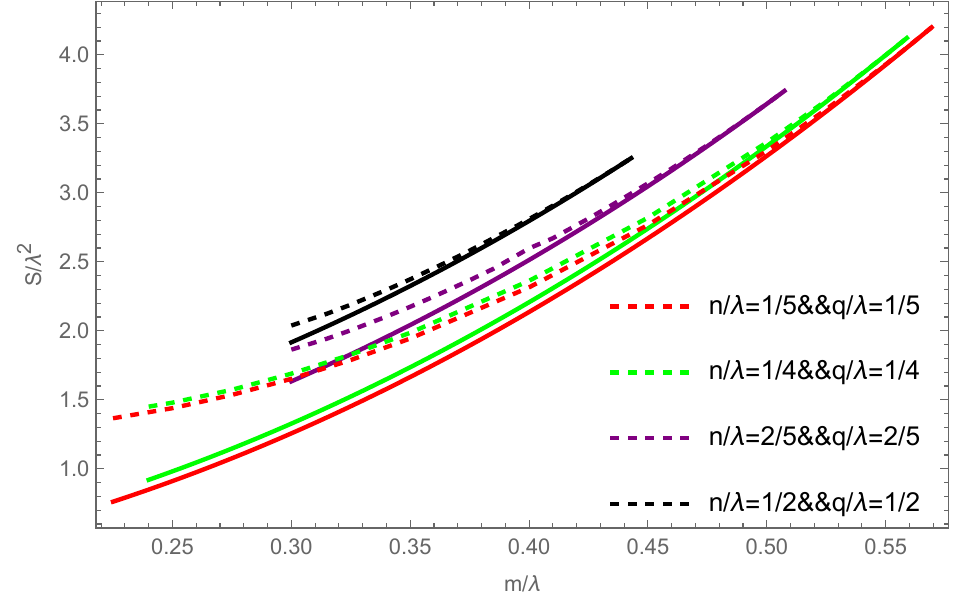}
	}
	\\
	\subfloat[$T-m$ curve for fixed $n/\lambda=1/4$]{
		\centering
		\includegraphics[width=0.5\linewidth]{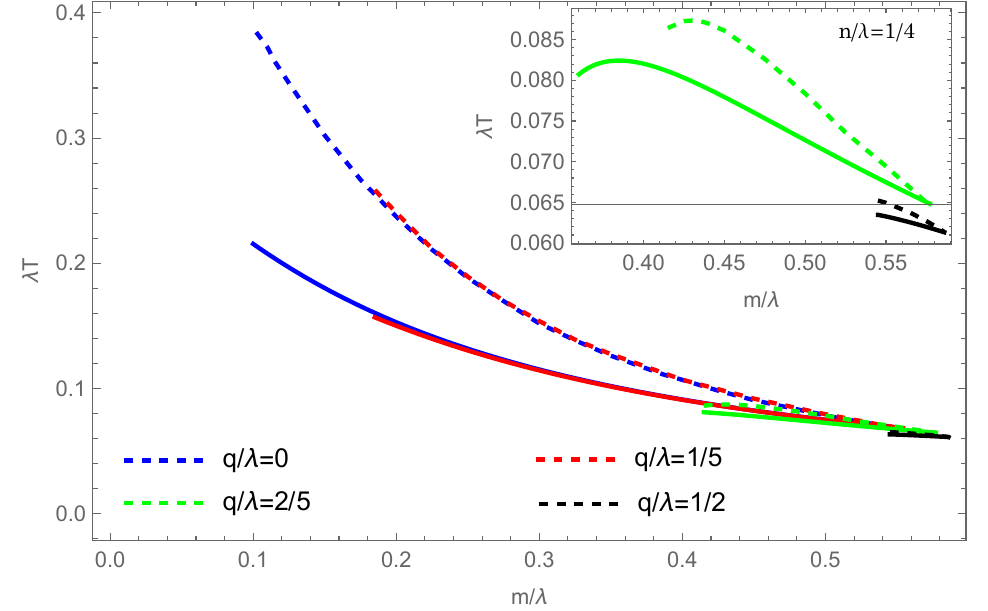}	
	}
	\subfloat[$T-m$ curve for for  $n=q$]{
		\includegraphics[width=0.5\linewidth]{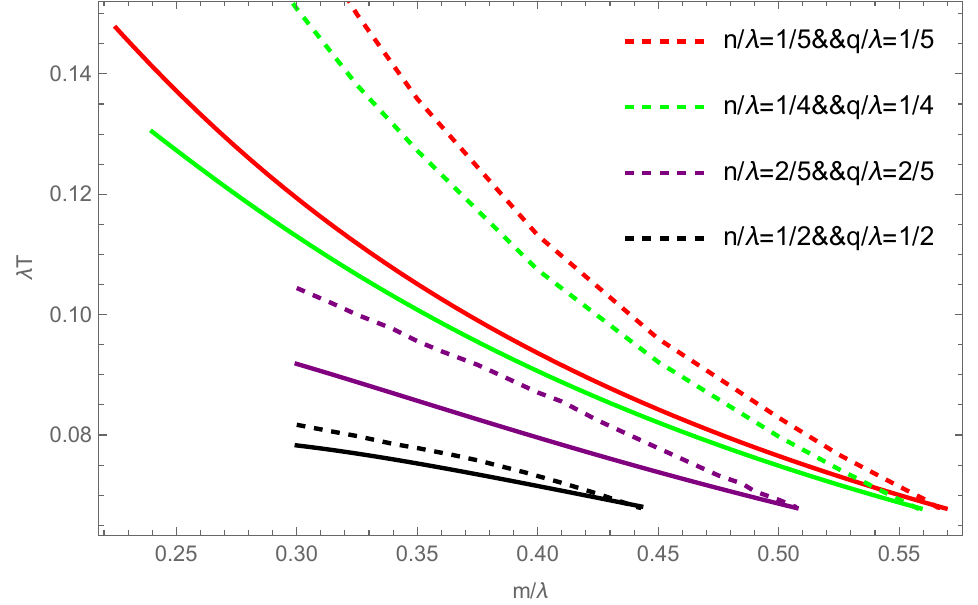}
	}
	\caption{$S/\lambda^2$ and $\lambda T$ as functions of $m/\lambda$ for fixed $n/\lambda=1/4$ and $n=q$ cases. The dashed and solid lines correspond to the charged scalarized black hole and the scalar-free Taub-NUT black hole, respectively.}
	
\end{figure}

Turning to the thermodynamic properties, we present the entropy and temperature as functions of the mass parameter for the two cases (fixed $n/\lambda=1/4$ and $n=q$) in Fig.10. One can see that the entropy of the scalarized black hole is still larger than that of the scalar-free black holes for both cases. However, the phenomenon that the maximum entropy of hairy black holes at the bifurcation points is almost constant for a range of $m$ no longer emerges. Furthermore, as $q/\lambda$ increases, the entropy values decrease for both hairy black holes and analytic scalar-free charged Taub-NUT black holes. Another property observed in the $D$-$m$ curves—that increasing the charge shortens the hairy solution—can also be seen in the entropy curves. Additionally, for larger $q$, it is difficult to distinguish the hairy black holes from the scalar-free Taub-NUT solutions. As for the temperature, these $T(m)$ curves have a similar pattern to the fixed $q/\lambda$ case. First, the temperature of hairy black holes is larger than that of scalar-free Taub-NUT black holes. Second, increasing the charge will make the temperature relatively higher.

\section{Conclusion}
In this work, we investigate the scalarization of charged Taub-NUT black holes in the extended scalar-tensor-Gauss-Bonnet gravity theory in the presence of an electromagnetic field, where the non-trivial scalar hair is triggered by the curvature of the spacetime through the Gauss-Bonnet coupling term.  Specifically, we employ an exponential coupling function, which allows the theory to admit the analytic charged Taub-NUT black hole as a vacuum solution.To identify the bifurcation points where scalarization originates, we perform a linear perturbation analysis of the scalar field against the charged Taub-NUT background. Since the charged Taub-NUT solution is characterized two independent integration constants : NUT parameter $n$ and electric parameter $q$, the resulting parameter space is two-dimensional. We explore this space along three distinct trajectories: fixed $q/\lambda$, fixed $n/\lambda$, and the symmetric case where $n = q$. By numerically solving the scalar field equation (Eq. \ref{scalar equation}), we generate three panels of existence lines, which serve as a foundational guide for constructing fully back-reacted hairy charged Taub-NUT black hole solutions.

We conducted an intensive study of the case with a fixed electric parameter, $q/\lambda = 1/4$. The scalar field parameter $D$ is presented as a function of the event horizon radius $r_h$ and the mass parameter $m$ across various NUT parameters. From these results, we observe a distinct phenomenon: within a moderate range of the NUT parameter, the hairy black hole solutions exhibit a dual-branch structure. Conversely, for sufficiently large or small values of $n$, the numerical solutions collapse into a single branch. In the remaining two cases—fixed $n$ and the symmetric case $n=q$—we observe a consistent effect of the electric charge on the hairy solutions. An increase in electric charge leads to the shortening of solution curves, and this effect is more clearly demonstrated by the entropy curve presented in Fig. 8(a).

We further investigated the thermodynamic properties of these hairy black holes, specifically focusing on their temperature and entropy. Across all examined cases, the temperature of the scalarized black holes consistently exceeds that of their scalar-free Taub-NUT counterparts. The entropy, however, exhibits a more complex phenomenology compared to the temperature. We identify two universal characteristics across the three parameter regimes (fixed $q/\lambda$, fixed $n/\lambda$, and $n=q$): first, the entropy of the hairy solution is strictly greater than that of the scalar-free solution; second, the entropy reaches a  maximum precisely at the bifurcation point. Beyond these commonalities, we observe a novel feature unique to the fixed charge regime: the maximum entropy at the bifurcation point remains universal over a specific range of the mass parameter. Furthermore, increasing the electric charge $q$  narrows this mass range while simultaneously elevating the maximum entropy value.

The underlying physics responsible for this universal entropy bound deserves further exploration. Conserved charges such as mass and electric charge are generally difficult to compute for black holes in higher-derivative gravity theories, and this task becomes even more challenging for the numerical solutions obtained in the present work. We hope to make progress along this direction in the near future.

\section*{Acknowledgement}
This work is supported in part by NSFC (National Natural Science Foundation of China) Grants No.~12575061, Tianjin University Graduate Liberal Arts and Sciences Innovation Award Program (2023) No. B1- 2023-005 and Tianjin University Self-Innovation Fund Extreme Basic Research Project Grant No. 2025XJ21-0007.

\end{document}